\documentstyle[12pt]{article}

\begin{document}


\begin{center}
{\large {\bf {Motion in Brane World Models: The Bazanski Approach}}}

{\bf {M.E.Kahil{\footnote{{\bf Mathematics Department, Modern Sciences and
Arts University, Giza, EGYPT\newline
e.mail: kahil@aucegypt.edu}}}} }
\end{center}

\abstract {Recently, path equations have been obtained for
charged, spinning objects in brane world models, using a modified
Bazanski Lagrangian. In this study, path deviation equations of
extended objects are derived. The significance of moving extended
objects in brane world models is examined. Motion in non-
symmetric brane world models is also considered.}

\subsection*{The Bazanski Approach in Riemannian Geometry}

Geodesic and geodesic deviation equations can be obtained simultaneously by
applying the action principle on the  Bazanski Lagrangian [1]:
\begin{equation}
L= g_{\alpha \beta} U^{\alpha} \frac{D \Psi^{\beta}}{Ds},
\end{equation}
where $\frac{D}{Ds}$ is the covariant derivative. By taking the
variation with respect to the deviation vector ${\Psi^{\rho}}$ one
obtains the geodesic equation. Taking the variation with respect
to the unit tangent vector $U^{\rho}$, one obtains its geodesic
deviation equation respectively :
\begin{equation}
\frac{dU^{\alpha}}{ds} +{{\alpha} \brace {\mu \nu}}U^{\mu}U^{\nu}=0 ,
\end{equation}
\begin{equation}
\frac{D^{2}\Psi^{\alpha}}{Ds^{2}} = R^{\alpha}_{. \beta \gamma \delta}
U^{\beta}U^{\gamma} \Psi^{\delta},
\end{equation}
where ${{\alpha} \brace {\mu \nu}}$ is the Christoffel symbol of the second
kind $R^{\alpha}_{\beta \gamma \delta}$ is the Riemann- Christoffel
curvature tensor.

It is worth mentioning that the Bazanski approach has been
successfully applied in geometries different from the Riemannian
one [2],[3]. Also, the Lagrangian (1) can be amended to describe
path and path deviation equations of spinning charged particles
[4] by introducing the following Lagrangian:
\begin{equation}
L = g_{\alpha \beta} U^{\alpha}{\frac{D \Psi^{\beta}}{Ds}} + ( \frac {e}{m}
F_{\alpha \beta} U^{\beta} + \frac{1}{2m} R_{\alpha \beta \gamma \delta}
S^{\gamma \delta} U^{\alpha} ) \Psi^{\mu}
\end{equation}
to give
\begin{equation}
\frac{dU^{\alpha}}{d s}+{{\alpha} \brace {\mu \nu}}U^{\mu}U^{\nu}= \frac{e}{m%
}F^{\mu}_{. \nu} U^{\nu}+\frac{1}{2m} R^{\alpha}_{. \mu \nu \rho}
S^{ \nu \rho} U^{\mu}.
\end{equation}
Its spinning charged deviation equation becomes:
\[
\frac{D^{2}\Psi^{\alpha}}{Ds^{2}}= R^{\alpha}_{.\mu
\nu\rho}U^{\mu}U^{\nu}\Psi^{\rho} +\frac{e}{m}(F^{\alpha}_{.\nu} \frac{D
\Psi^{\nu}}{Ds}+F^{\alpha}_{.\nu ; \rho}U^{\nu}\Psi^{\rho})
\]
\begin{equation}
~~~~~~~~~~~~~~ +{\frac{1}{2m}} ( R^{\alpha}_{. \mu \nu \rho} S^{\nu \rho}
\frac{D \Psi^{\nu}}{Ds} + R^{\alpha}_{\mu \nu \lambda}S^{\nu \lambda}_{.;
\rho}U^{\mu}\Psi^{\rho} + R^{\alpha}_{\mu \nu \lambda; \rho }S^{\nu \lambda}
U^{\mu} \Psi^{\rho}),
\end{equation}
where $F^{\mu}_{.\nu}$ is the electromagnetic field tensor and
$S^{\gamma \delta}$ is the spin tensor .

\subsection*{The Bazanski Approach in Brane World Models}

It is well known that in the Brane world scenario our universe can
be described in terms of a 4+N dimensional, with $N \geq 1$ and the
4D space-time part of it is embedded in a 4+N manifold [5].
Accordingly, the bulk geodesic motion is observed by a four
dimensional observer to reproduce the physics of 4D space-time
[6]. Consequently, it is vital to derive the path and the path
deviation equations for a test particle on a brane using the
following Lagrangian [7] :
\begin{equation}
L = g_{\mu \nu}(x^{\rho},y) U^{\mu} \frac{D \Psi ^{\nu}}{Ds}
+f_{\mu}\Psi^{\mu},
\end{equation}
where $g_{\mu \nu}(x^{\rho},y)$ is the induced metric and$f_{\mu} = \frac{1}{%
2} U^{\rho}U^{\sigma} \frac{\partial g_{\rho \sigma}}{\partial y}\frac{dy}{ds%
} U_{\mu}$ describes a parallel force due to the effect of
non-compactified extra dimension. The variation of the Lagrangian
gives [8]:
\begin{equation}
\frac{dU^{\mu}}{ds}+ {{\mu} \brace {\alpha \beta}} {U ^{\alpha}}{U^{\beta}}
= (\frac{1}{2}U^{\rho}U^{\sigma}-g^{\rho \sigma}) \frac{\partial g_{\rho
\sigma}}{\partial y}\frac{dy}{ds} U^{\mu} .
\end{equation}
As in the brane world models, one can express $\frac{1}{2}
\frac{\partial
g_{\rho \sigma}}{\partial y}$ in terms of the extrinsic curvature $%
\Omega_{\rho \sigma}$ i.e. $\Omega_{\alpha \beta} = \frac{1}{2} \frac{%
\partial g_{\rho \sigma}}{\partial y} $ [9]. Thus, equation (8) becomes:
\begin{equation}
\frac{dU^{\mu}}{ds}+ {{\mu} \brace {\alpha \beta}} {U ^{\alpha}}{U^{\beta}}
=2(\frac{1}{2}U^{\mu}U^{\sigma}-g^{\mu \sigma}) \Omega_{\rho \sigma} \frac{dy%
}{ds} U^{\rho}.
\end{equation}
And its corresponding deviation equation is
\begin{equation}
\frac{D^{2}\Psi^{\alpha}}{Ds^{2}}= R^{\alpha}_{.\mu
\nu\rho}U^{\mu}U^{\nu}\Psi^{\rho}
+(U^{\alpha}U^{\sigma}U^{\nu})\Omega_{\sigma \nu}{\partial y}\frac{dy}{ds}%
)_{; \rho} \Psi^{\rho} + 2(( \frac{1}{2}U^{\alpha}U^{\sigma}-g^{\alpha
\sigma})\Omega_{\rho \sigma}\frac{dy}{ds}U^{\rho})_{; \delta} \Psi^{\delta}
\end{equation}
Also, applying the bazanski approach in brane world models, we obtain the path and path
deviation equations for a spinning charged object, respectively
\begin{equation}
\frac{dU^{\alpha}}{ds} +{{\alpha} \brace {\mu \nu}}U^{\mu}U^{\nu}= \frac{e}{m%
}F^{\alpha}_{\nu}U^{\nu} + \frac{1}{2m}R^{\alpha}_{\beta \mu \nu} S^{\mu
\nu}U^{\beta} + 2(\frac{1}{2}U^{\alpha}U^{\rho} -g^{\alpha} \rho)
\Omega_{\rho \delta}\frac{dy}{ds} U^{\delta}
\end{equation}

and
\[
\frac{D^{2}\Psi^{\alpha}}{Ds^{2}}= R^{\alpha}_{.\mu
\nu\rho}U^{\mu}U^{\nu}\Psi^{\rho} +\frac{1}{2m}( R^{\alpha}_{. \mu \nu \rho}
S^{\nu \rho} \frac{D \Psi^{\mu}}{Ds}+ R^{\alpha}_{\mu \nu \lambda}S^{\nu
\lambda}_{.; \rho}U^{\mu}\Psi^{\rho} + R^{\alpha}_{\mu \nu \lambda; \rho
}S^{\nu \lambda} U^{\mu} \Psi^{\rho})
\]
\begin{equation}
~~~+\frac{e}{m} ( F^{\alpha}_{\rho}\Psi^{\rho} + F^{\alpha}_{\rho} \frac{D
\Psi^{\rho}}{Ds}) +(\frac{1}{2}U^{\alpha}U^{\sigma}U^{\nu}) \Omega_{\sigma
\nu }\frac{dy}{ds})_{; \rho} \Psi^{\rho} + (\frac{1}{2}U^{\alpha}U^{%
\sigma}-g^{\alpha \sigma})\Omega_{\sigma \nu}\frac{dy}{ds}\frac{D \Psi^{\nu}%
}{Ds}.
\end{equation}
Thus, equations (11) and (12) are derived from the following Lagrangian
\begin{equation}
L= g_{\mu \nu}(x^{\rho},y) U^{\mu} \frac{D \Psi^{\nu}}{Ds} + 2(\frac{1}{2m}
R_{\mu \nu \rho \sigma} S^{\rho \sigma} U^{\nu} + \Omega_{\rho \sigma }{%
\partial y} U^{\rho}U^{\sigma} U_{\mu} \frac{dy}{ds} ) \Psi ^{\mu},
\end{equation}

\subsection*{ Path \& Path Deviation Equations of Non-Symmetric Geometries
in Brane World Models}

Path equation and Path deviation equations in Brane World Models defined in non-symmetric
geometries can be obtained by suggesting the following Lagrangian: 
\begin{equation}
L = {\bf {g}}_{\mu \nu} U^{\mu} \frac{D \Psi^{\nu}}{D \tau} + \lambda
f_{[\mu \nu]}U^{\mu}\Psi^{\nu}+ \frac{1}{2}U^{\alpha}U^{\beta}U^{\rho} \frac{%
\partial {\bf {g}_{\alpha \beta}}}{\partial s} \frac{dy}{ds},
\end{equation}
where ${\bf {g}}_{\mu \nu} = g_{(\mu \nu)} + g_{[\mu \nu]} $,$\lambda$ is a
parameter and, $f_{[ \mu \nu] }= \hat{A}_{\mu ,\nu} - \hat{A}_{\nu , \mu}$
is a skew symmetric tensor related to the Yukawa force [10].

Applying the Bazanski approach we obtain the path equation
\begin{equation}
\frac{dU^{\alpha}}{ds} +{{\alpha} \brace {\mu \nu}}U^{\mu}U^{\nu}= \lambda
{\bf {g}}^{ \alpha \mu } f_{[ \mu \nu ]} U^{\nu} + {\bf {g}}^{\alpha \sigma
} g_{[ \nu \sigma ]; \rho} U^{\nu}U^{\rho} +(\frac{1}{2}U^{\rho}U^{\sigma}-%
{\bf {g}}^{\rho \sigma}) \frac{\partial {\bf {g}}_{\rho \sigma}}{\partial y}%
\frac{dy}{ds} U^{\mu} .
\end{equation}
and its path deviation equation:
\[
\frac{D^{2}\Psi^{\alpha}}{Ds^{2}}= R^{\alpha}_{.\mu
\nu\rho}U^{\mu}U^{\nu}\Psi^{\rho}+2 {\bf {g}}^{\sigma \alpha} ( g_{[\nu [
\sigma];\rho] } ) \frac{D \Psi^{\nu}}{Ds}U^{\rho} +\lambda(f^{\alpha}_{.\nu}
\frac{D \Psi^{\nu}}{Ds}+f^{\alpha}_{.\nu ; \rho}U^{\nu}\Psi^{\rho}).
\]
\begin{equation}
~~~+ ((\frac{1}{2}U^{\alpha}U^{\rho}-{\bf {g}}^{\alpha \rho})\frac{\partial
{\bf {g}}_{\delta \rho}}{dy}U^{\delta} \frac{dy}{ds}) _{;\nu} \Psi^{\nu} + (%
\frac{1}{2}U^{\alpha}U^{\mu}\frac{\partial{\bf {g}}_{\mu \nu }}{dy} \frac{dy%
}{ds})\frac{D \Psi^{\nu}}{Ds}
\end{equation}

\subsection*{The Bazanski Approach in Curved Clifford Space}

It is well known that extended objects can be expressed by
p-branes[11]. This type of representation is defined in curved
Clifford Space. The advantage of this space is to show that
extended objects are purely poly-geodesics satisfying a
metamorphic transformation.[12].

Thus, we suggest the following Lagrangian:
\begin{equation}
L=g_{AB} P^{\alpha} \frac{\nabla \phi^{\beta}}{\nabla \tau} + \frac{1}{2}
S_{\alpha \beta}\frac{\nabla \phi^{\alpha \beta} }{\nabla \tau
},
\end{equation}
such that
\[
\frac{\nabla \phi^{\alpha}}{\nabla \tau} = \frac{d \phi^{\alpha}}{d \tau} +
\Gamma ^{\alpha}_{\beta \rho} U^{\rho} \phi^{\beta} + \frac{1}{2m}{\hat{R}}%
^{\alpha}_{. \beta \gamma \delta}S^{\gamma \delta} \phi^{\beta} + \frac{1}{2m%
}S^{\omega}_{\gamma}(\Xi_{\mu \omega}^{~~\alpha \sigma}U^{\mu} + \frac{1}{2m}%
S^{\rho \beta} \Omega_{\rho \beta \omega}^{~~ \alpha \sigma} ),
\]
where $\Gamma^{\alpha}_{\beta \rho}$ is the Cartan connection, ${\hat{R}}%
^{\alpha}_{. \beta \gamma \delta}$ is the Cartan curvature and the
two other metamorphic connections $\Xi_{\beta \rho}^{~~\alpha
\delta} \& \Omega{\beta \rho \gamma}^{~~\alpha \delta} $ existed
due to the presence of bivector
quantities. By taking variation with respect to its deviation vector $%
\phi^{\delta}$ we get
\begin{equation}
\frac{\nabla U^{\alpha}}{\nabla \tau} =0
\end{equation}
and taking the variation with respect to its deviation bivector $%
\phi^{\delta \rho}$ we obtain
\begin{equation}
\frac{\nabla S^{\alpha \beta}}{\nabla \tau} =0
\end{equation}
In the case of a charged particle described by a polyvector we can
find that the corresponding linear momentum equation becomes [12]
\begin{equation}
\frac{\nabla U^{\alpha}}{\nabla \tau} =\frac{e}{m} F^{\alpha}_{\beta}U^{%
\beta}
\end{equation}
and its angular momentum equation takes the following form
\begin{equation}
\frac{\nabla S^{\alpha \beta}}{\nabla \tau} = F^{\alpha}_{\nu}S^{\nu \beta}-
F^{\beta}_{\nu}S^{\nu \alpha}
\end{equation}
We suggest the following Lagrangian to derive (20) and (21)
\begin{equation}
L=g_{\mu \nu} P^{\mu} \frac{\nabla \phi^{\nu}}{\nabla \tau } + \frac{1}{2}S_{\mu \nu} \frac{%
\tau\phi^{ \mu \nu}}{\nabla \tau } + F_{\mu \nu} U^{\nu} \phi^{\mu} + \frac{1}{2}(F_{\mu
\rho}S^{\rho}_{\nu}-F_{\nu}^{~\rho}S_{\rho \mu} )\phi^{\mu \nu}.
\end{equation}
If we put the flavor of Kaluza-Klein in curved Clifford space for
combining gravity and electromagnetism, i.e, increasing the
spatial dimension by a compacted one, then equations (20),(21)  can be
derived from the following Lagrangian:
\begin{equation}
L=g_{AB} P^{A} \frac{\nabla \phi^{B}}{\nabla \hat \tau} + \frac{1}{2} S_{AB} \frac{\nabla \phi^{ AB}}{\nabla
\hat \tau },
\end{equation}
where $A= 1,2,3,4,5$ to become
\begin{equation}
\frac{\nabla U^{\alpha}}{\nabla \hat \tau} =0
\end{equation}
and
\begin{equation}
\frac{\nabla S^{\alpha \beta}}{\nabla \hat \tau} =0
\end{equation}
which means that the path of a charged spinning particle defined
in Riemannain geometry behave like as a test particle in 5D curved
Clifford space.

\subsection*{Discussion and Concluding Remarks}

In this study, we have derived path and path deviation equations
for test particles and spinning charged test objects in Brane World Models from
one single Lagrangian. Also, the procedure has been used to derive
path and path deviation equations for a test particle existing in
non-symmetric theory of gravity and examined how the extrinsic
curvature term would be amended due to the existence of
non-symmetric terms of gravitational field. Thus, we can find that the added term appeared in the extrinsic
curvature of non-symmetric part may be connected to the spin.
Finally, we have dealt with extended objects as p-branes defined
in curved Clifford space [13]. This type of space has a deeper
understanding of physics, i.e., quantities in nature could be
defined by polyvectors. From this perspective we have developed
its corresponding Bazanski Lagrangian to include deviation
bi-vectors together with deviation vectors. The importance of this
approach is that we can derive from one Lagrangian(17) two
simultaneous quantities responsible for conservation of momentum
(18) and angular momentum (19) respectively. Using this mechanism,
we have found that the Dixon-Souriau equations in Riemannian
geometry could be seen like equations for charged object and their
angular momentum part defined in curved clifford spaces. In
addition, if we have the flavor of Kaluza-Klein of unifying gravity
and electromagnetism in curved Clifford space, we must increase
the dimension of this space by an extra compacted dimension to
preserve the conservation of charges. Consequently, from equations
(24) and (25) we can find that charged spinning particles behave
like test particles defined in higher dimensional curved clifford
space. To conclude this study, we must take into consideration that
unification processes can be achieved if we increase the number of
dimensions and extend the geometries to geometrize all quantities
that appeared in our approach. \newline {\bf
{Acknowlgements}}\newline The author would like to thank
Professors M.I. Wanas, M. Abdel-Megied , G.S. Hall, T. Harko,  G.
De Young and Mr. W.S. El-Hanafy for their useful comments. A word of thanks should be
addressed to Professors A. Rajantie and F.H. Stoica for their suppport and
help to pariticapate in PASCOS07. A very special word of
thankfulness should be sent to Dr. N. El-Degwi and Professor K. Abdel
Hamid for helping me to get a grant to represent MSA University at
this confenence.\\
{\bf {References}} \newline
{[1]} Bazanski, S.L. (1989) J. Math. Phys., {\bf {30}}, 1018. \newline
{[2]} Wanas, M.I., Melek, M. and Kahil, M.E.(1995) Astrophys. Space Sci.,%
{\bf {228}}, 273.; 

gr-qc/0207113. \newline
{[3]} Wanas, M.I. and Kahil, M.E.(1999) Gen. Rel. Grav., {\bf {31}}, 1921. ;
gr-qc/9912007 \newline
{[4]} Kahil, M.E. (2006), J. Math. Physics {\bf {47}},052501. \newline
{[5]} Liu, H. and Mashhoon, B. (2000), Phys. Lett. A, {\bf {272}},26 ;
gr-qc/0005079 \newline
{[6]} Youm, D. (2000) Phys.Rev.D{\bf {62}}, 084002; hep-th/0004144 \newline
{[7]} Kahil, M.E. (2006) a paper presented at the Eleventh meeting of 

Marcel Grosmann, Berlin 23-30 July 2006; gr-qc/0701015 \newline
{[8]}Ponce de Leon, J. (2001) Phys Lett B, {\bf {523}} ;gr-qc/0110063
\newline
{[9]}Dick, R. Class. Quant. Grav.(2001), {\bf {18}}, R1. \newline
{[10]} Legar\'{e}, J. and Moffat, J.W. (1996), Gen. Rel. Grav., {\bf {26}},
1221.\newline
{[11]}Castro, C. and Pavsic, M. (2002) Phys. Lett B {\bf {539}}, 133;
hep-th/0110079 \newline
{[12]} Pizzaglia, W.M. (1999) gr-qc/9912025 \newline
{[13]}Castro,C. (2002) physics/0011040 \newline

\end{document}